\documentclass[a4paper, 12pt]{article}

\usepackage{bm}
\usepackage{amssymb,amsthm,amsmath}
\usepackage{color}
\usepackage{natbib}
\renewcommand{\d}{{\rm d}}
\newcommand{\Card}{\rm Card}
\newcommand{\sgn}{{\rm sgn}}

\title{Linear regression in the Bayesian framework}
\author{Thierry Alex Mara\footnote{{University of Reunion, thierry.mara(at)univ-reunion.fr}}}
\date{}

\begin{document}

\maketitle

\noindent{\bf Keywords}: linear regression, Bayesian inference, generalized least-squares, ridge regression, LASSO, model selection criterion, KIC

\begin{abstract}
These notes aim at clarifying different strategies to perform linear regression from given dataset. Methods like the weighted and ordinary least squares, ridge regression or LASSO are proposed in the literature. The present article is my understanding of these methods which are, according to me, better unified in the Bayesian framework. The formulas to address linear regression with these methods are derived. The KIC for model selection is also derived in the end of the document.
\end{abstract}
\newpage

\section{Introduction}
Let's consider an input-output relationship of the form:
\begin{equation}
\label{Eq:Model1}
y = \sum_{\bm{\alpha}\in\mathcal{A}}a_{\bm{\alpha}}\psi_{\bm{\alpha}}(\bm{x}) + \epsilon
\end{equation}
where $\bm{\alpha}=\alpha_1\dots\alpha_d$ is a multi-dimensional index, $\mathcal{A}\subset\mathbb{N}^d$ is a finite subset of multi-indexes and $\psi_{\bm{\alpha}}(\bm{x})$ is some function independent of $x_i$ if $\alpha_i=0$, and $\epsilon$ an error term. Given the sets of observations $\bm{y}=\left\lbrace y_{n}\right\rbrace_{n=1}^{N}$ and $\bm{x}=\left\lbrace x_{n1},\dots,x_{nd}\right\rbrace_{n=1}^{N}$, the matrix form of \eqref{Eq:Model1} is,
\begin{equation}
\label{Eq:Model1_Matrix}
\bm{y} = \bm{\Psi}\bm{a} + \bm{\epsilon}.
\end{equation}
The error vector $\bm{\epsilon}$ accounts for i) possible \emph{stochastic} error $\epsilon_s$ (e.g. measurement noise, stochastic model noise), ii) sampling error $\epsilon_d$ (as $(\bm{y},\bm{x})$ is just one possible sample), and iii) truncation error $\epsilon_t$ (because ${\Card}(\mathcal{A})$ is finite). We write $\epsilon=\epsilon_s+\epsilon_d+\epsilon_t$. We note that when one has to deal with $y$ as the response of a \emph{deterministic} mathematical model, stochastic errors $\epsilon_s=0$. Sampling error can be assessed with several samples $(\bm{y},\bm{x})$. In practice, this is hardly possible, and one can referred to the \emph{bootstrap} technique to assess $\epsilon_d$ \citep{Efron93CHAP}. Both $\epsilon_d$ and $\epsilon_t$ are reducible by increasing the number of observations $N$ (and the diversity of the observations of course).

The first issue addressed in this short note is: \emph{how can we infer $\bm{a}=\left\lbrace a_{\bm{\alpha}}: \bm{\alpha}\in\mathcal{A}\right\rbrace$, given $(\bm{y},\bm{x})$ and $\mathcal{A}$?} The second one is \emph{how to infer the best subset $\mathcal{A}$ given $(\bm{y},\bm{x})$?}. These issues are addressed in a Bayesian framework which requires some assumptions about the error's probability density function (called the \emph{likelihood function}) and some information about the unknown coefficients themselves (called \emph{prior} belief). Having assumed the likelihood is $\epsilon\sim p_{\epsilon}(\epsilon)$, we can also write that $\left(y-\sum_{\bm{\alpha}\in\mathcal{A}}a_{\bm{\alpha}}\psi_{\bm{\alpha}}(\bm{x})\right)\sim p_{\epsilon}\left(y-\sum_{\bm{\alpha}\in\mathcal{A}}a_{\bm{\alpha}}\psi_{\bm{\alpha}}(\bm{x})\right)$ which is usually written: $p_{\bm{y}\vert \bm{a},\mathcal{A}}=p(\bm{y}\vert\bm{a},\mathcal{A})$ or simply $\mathcal{L}(\bm{y}\vert\bm{a},\mathcal{A})$ (the $\mathcal{L}$ standing for likelihood). The prior is denoted $p_{\bm{a}\vert\mathcal{A}}=p(\bm{a}\vert\mathcal{A})$.

The first issue is addressed by assessing the coefficient joint posterior density function (pdf inferred from Bayes rule), namely
\begin{equation}
\label{Eq:Posterior}
p(\bm{a}\vert\bm{y},\mathcal{A}) = \frac{p(y\vert \bm{a},\mathcal{A})p(\bm{a}\vert \mathcal{A})}{p(\bm{y}\vert\mathcal{A})}.
\end{equation}
Because \eqref{Eq:Model1} is linear w.r.t. the vector of coefficients $\bm{a}$, straightforward calculations can be employed as opposed to more demanding approach like Markov chains Monte Carlo. These straightforward calculations are derived in the present document.

The second issue can be addressed by evaluating the Bayesian Model Evidence (BME) defined as:
\begin{equation}
\label{Eq:BME}
p(\bm{y}\vert\mathcal{A}) =\int_{\mathbb{R}^P}p(\bm{y},\bm{a}\vert \mathcal{A})\d\bm{a}
\end{equation}
where $P=\Card(\mathcal{A})$ is the number of coefficients. Indeed, Eq.\eqref{Eq:BME} represents how well the current model $(\bm{a},\mathcal{A})$ explains (fits) the observed data $\bm{y}$. Finding the subset $\mathcal{A}$ that maximizes \eqref{Eq:BME} provides the answer to the second question.
\section{Gaussian Likelihood}
\label{Sec:GLS}
Let us assume that $p_{\bm{\epsilon}}\sim\mathcal{N}(\bm{0},\bm{C}_{M})$, this leads to the following Gaussian likelihood:
\begin{equation}
\label{Eq:Gaussian_Likelihood}
p(\bm{y}\vert\bm{a},\mathcal{A})=(2\pi)^{-N/2}\vert\bm{C}_M\vert^{-1/2}\exp\left(-\frac{1}{2}\left(\bm{y}-\bm{\Psi}\bm{a}\right)^t\bm{C}_M^{-1}\left(\bm{y}-\bm{\Psi}\bm{a}\right)\right)
\end{equation}
where $\vert\cdot\vert$ stands for the determinant.

\subsection{Uniform Prior \& Maximum Likelihood Estimate}
\label{Sec:MLE}
Let us assume further that the prior knowledge about the coefficient values is a rectangular domain $\Omega=[l_1,u_1]\times\dots\times[l_d,u_d]$. This means that $p_{\bm{a}\vert\mathcal{A}}=cste$ if $\bm{a}\in\Omega$, and zero elsewhere. As a consequence, the joint pdf of the coefficients is,
\begin{equation}
\label{Eq:Posterior1}
p(\bm{a}\vert\bm{y},\mathcal{A}) \propto p(\bm{y}\vert \bm{a},\mathcal{A})\propto \exp\left(-\frac{1}{2}\left(\bm{y}-\bm{\Psi}\bm{a}\right)^t\bm{C}_M^{-1}\left(\bm{y}-\bm{\Psi}\bm{a}\right)\right), \bm{a}\in\Omega
\end{equation}

\noindent\underline{Analytical Solution}: The term in the exponential in \eqref{Eq:Posterior1},
$$
\left(\bm{y}-\bm{\Psi}\bm{a}\right)^t\bm{C}_M^{-1}\left(\bm{y}-\bm{\Psi}\bm{a}\right)=
\underset{cste/\bm{a}}{\underbrace{\bm{y}^t\bm{C}_M^{-1}\bm{y}}}
-\underset{=\bm{a}^t\bm{\Psi}^t\bm{C}_M^{-1}\bm{y}}{\underbrace{\bm{y}^t\bm{C}_M^{-1}\bm{\Psi}\bm{a}}}
-\bm{a}^t\bm{\Psi}^t\bm{C}_M^{-1}\bm{y}
+\bm{a}^t\bm{\Psi}^t\bm{C}_M^{-1}\bm{\Psi}\bm{a}
$$
the second underbrace true because $\bm{C}_M$ is symmetric (and positive-definite) by definition. So, we get
$$
\left(\bm{y}-\bm{\Psi}\bm{a}\right)^t\bm{C}_M^{-1}\left(\bm{y}-\bm{\Psi}\bm{a}\right)=
\bm{a}^t\bm{\Psi}^t\bm{C}_M^{-1}\bm{\Psi}\bm{a} - 2\bm{a}^t\bm{\Psi}^t\bm{C}_M^{-1}\bm{y} + c
$$
\begin{small}
$$
=\left(\left[\bm{\Psi}^t\bm{C}_M^{-1}\bm{\Psi}\right]^{1/2}\bm{a} - \left[\bm{\Psi}^t\bm{C}_M^{-1}\bm{\Psi}\right]^{-1/2}\bm{\Psi}^t\bm{C}_M^{-1}\bm{y}\right)^t\left(\left[\bm{\Psi}^t\bm{C}_M^{-1}\bm{\Psi}\right]^{1/2}\bm{a} - \left[\bm{\Psi}^t\bm{C}_M^{-1}\bm{\Psi}\right]^{-1/2}\bm{\Psi}^t\bm{C}_M^{-1}\bm{y}\right) + c
$$
\end{small}
Factorizing by $\left[\bm{\Psi}^t\bm{C}_M^{-1}\bm{\Psi}\right]^{1/2}$ in both parentheses yields,
\begin{small}
$$
\left(\bm{y}-\bm{\Psi}\bm{a}\right)^t\bm{C}_M^{-1}\left(\bm{y}-\bm{\Psi}\bm{a}\right) =
\left(\bm{a} - \left[\bm{\Psi}^t\bm{C}_M^{-1}\bm{\Psi}\right]^{-1}\bm{\Psi}^t\bm{C}_M^{-1}\bm{y}\right)^t\bm{\Psi}^t\bm{C}_M^{-1}\bm{\Psi}\left(\bm{a} - \left[\bm{\Psi}^t\bm{C}_M^{-1}\bm{\Psi}\right]^{-1}\bm{\Psi}^t\bm{C}_M^{-1}\bm{y}\right) + C
$$
\end{small}
Replacing this result in \eqref{Eq:Posterior1} yields,
\begin{equation}
\label{Eq:Posterior_likelihood1}
\boxed{p(\bm{a}\vert\bm{y},\mathcal{A})=(2\pi)^{-P/2}\vert\bm{\tilde{C}}_{aa}\vert^{-1/2}\exp\left(-\frac{1}{2}\left(\bm{a}-\bm{\tilde{a}}\right)^t\bm{\tilde{C}}_{aa}^{-1}\left(\bm{a}-\bm{\tilde{a}}\right)\right)}
\end{equation}
where\\
$\bm{\tilde{a}}=\left[\bm{\Psi}^t\bm{C}_M^{-1}\bm{\Psi}\right]^{-1}\bm{\Psi}^t\bm{C}_M^{-1}\bm{y}$ is called the Maximum Likelihood Estimate (MLE)\\
$\bm{\tilde{C}}_{aa}=\left[\bm{\Psi}^t\bm{C}_M^{-1}\bm{\Psi}\right]^{-1}$ is the covariance associated to $\bm{a}$
and $P=\Card(\mathcal{A})$ is the number of coefficients.\\

\noindent\underline{N.B.}: Because the coefficients have been constrained within a finite rectangular domain $\Omega$, and $\bm{C}_M$ is assumed given, the posterior joint pdf should be written $p(\bm{a}\vert\bm{y},\mathcal{A},\bm{C}_M)=\mathcal{N}(\bm{\tilde{a}},\bm{\tilde{C}}_{aa})$ if $\bm{\tilde{a}}\in\Omega$ and $p(\bm{a}\vert\bm{y},\mathcal{A},\bm{C}_M)=0$ elsewhere.
\subsection{Homoscedastic Gaussian Error \& Ordinary Least Squares}
\label{Sec:OLS}
Assuming homoscedastic Gaussian error, that is, setting $\bm{C}_{M}=\sigma^2_{\epsilon}\bm{I}_N$ where $\bm{I}_N$ stands for $N\times N$ identity matrix, yields\\
$p(\bm{a}\vert\bm{y},\mathcal{A},\sigma^2_{\epsilon})=\mathcal{N}(\bm{\tilde{a}},\bm{\tilde{C}}_{aa})$\\
where\\
$\bm{\tilde{a}}=\left[\sigma^{-2}_{\epsilon}\bm{\Psi}^t\bm{I}_N\bm{\Psi}\right]^{-1}\sigma_{\epsilon}^{-2}\bm{\Psi}^t\bm{y}=\left[\bm{\Psi}^t\bm{\Psi}\right]^{-1}\bm{\Psi}^t\bm{y}$ which is the Ordinary Least-Square estimator,\\
and $\bm{\tilde{C}}_{aa}=\sigma_{\epsilon}^{2}\left[\bm{\Psi}^t\bm{\Psi}\right]^{-1}$.\\

In practice $\sigma_{\epsilon}^{2}$ is unknown, and because \eqref{Eq:Gaussian_Likelihood} becomes in this case,
\begin{equation}
\label{Eq:OLS}
\boxed{p(\bm{y}\vert\bm{a},\mathcal{A},\sigma^2_{\epsilon})=(2\pi\sigma^2_{\epsilon})^{-N/2}\exp\left(-\frac{1}{2\sigma^2_{\epsilon}}\left(\bm{y}-\bm{\Psi}\bm{a}\right)^t\left(\bm{y}-\bm{\Psi}\bm{a}\right)\right)}
\end{equation}
$$
\Leftrightarrow -2\ln(p(\bm{y}\vert\bm{\tilde{a}},\mathcal{A},\sigma^2_{\epsilon}))=c+N\ln(\sigma^{2}_{\epsilon})+\frac{\left(\bm{y}-\bm{\Psi}\bm{\tilde{a}}\right)^t\left(\bm{y}-\bm{\Psi}\bm{\tilde{a}}\right)}{\sigma^{2}_{\epsilon}}
$$
we can infer that its MLE $\tilde{\sigma}^{2}_{\epsilon}$ is given by,
$$
-2\left.\frac{\d\ln(p(\bm{y}\vert\bm{\tilde{a}},\mathcal{A},\sigma^2_{\epsilon}))}{\d\sigma^2_{\epsilon}}\right\vert_{\sigma^2_{\epsilon}=\tilde{\sigma}^2_{\epsilon}}=0=\frac{N}{\tilde{\sigma}^{2}_{\epsilon}}-\frac{\left(\bm{y}-\bm{\Psi}\bm{\tilde{a}}\right)^t\left(\bm{y}-\bm{\Psi}\bm{\tilde{a}}\right)}{\tilde{\sigma}^{4}_{\epsilon}}
$$
$$
\Leftrightarrow \tilde{\sigma}^{2}_{\epsilon}=\frac{\left(\bm{y}-\bm{\Psi}\bm{\tilde{a}}\right)^t\left(\bm{y}-\bm{\Psi}\bm{\tilde{a}}\right)}{N}
$$

By noticing that \eqref{Eq:OLS} yields
\begin{equation}
p(\sigma^2_{\epsilon}\vert\bm{y},\bm{\tilde{a}},\mathcal{A})=(2\pi)^{-N/2}(\sigma^{-2}_{\epsilon})^{N/2}\exp\left(-\frac{\left(\bm{y}-\bm{\Psi}\bm{\tilde{a}}\right)^t\left(\bm{y}-\bm{\Psi}\bm{\tilde{a}}\right)}{2}\sigma^{-2}_{\epsilon}\right)
\end{equation}
we can infer that,
\begin{equation}
\boxed{p(\sigma^{-2}_{\epsilon}\vert\bm{y},\bm{\tilde{a}},\mathcal{A})\propto \left(\sigma^{-2}_{\epsilon}\right)^{k-1}\exp\left(-\frac{\sigma^{-2}_{\epsilon}}{\theta}\right)=\Gamma\left(\frac{N+2}{2},\frac{2}{\left(\bm{y}-\bm{\Psi}\bm{\tilde{a}}\right)^t\left(\bm{y}-\bm{\Psi}\bm{\tilde{a}}\right)}\right)}
\end{equation}
known as Gamma distribution whose mode is $\tilde{\sigma}^{2}_{\epsilon}$. Therefore, under homoscedastic Gaussian likelihood assumption the MLE of the posterior covariance is, $\bm{\tilde{C}}_{aa}=\tilde{\sigma}_{\epsilon}^{2}\left[\bm{\Psi}^t\bm{\Psi}\right]^{-1}$ with still $\bm{\tilde{a}}=\left[\bm{\Psi}^t\bm{\Psi}\right]^{-1}\bm{\Psi}^t\bm{y}$.
\subsection{Gaussian Prior \& Ridge Regression}
\label{Sec:Ridge_Regression}
Let us assume now that $p_{\bm{a}\vert\mathcal{A}}=\mathcal{N}(\bm{a}_0,\bm{C}_{aa})$. We note that $\Omega$ is no longer a rectangular domain but mostly a hyper-ellipsoid. Then, the joint posterior distribution of the coefficients is written,
\begin{equation}
p(\bm{a}\vert\bm{y},\mathcal{A}) \propto p(\bm{y}\vert \bm{a},\mathcal{A})p(\bm{a}\vert\mathcal{A})
\end{equation}
\begin{equation}
\label{Eq:Posterior2}
\Leftrightarrow p(\bm{a}\vert\bm{y},\mathcal{A})\propto \exp\left(-\frac{1}{2}\left[\left(\bm{y}-\bm{\Psi}\bm{a}\right)^t\bm{C}_M^{-1}\left(\bm{y}-\bm{\Psi}\bm{a}\right)+\left(\bm{a}-\bm{a}_0\right)^t\bm{C}_{aa}^{-1}\left(\bm{a}-\bm{a}_0\right)\right]\right)
\end{equation}

\noindent\underline{Analytical Solution}: Let us once again develop the term in the exponential, we get
$$
\left(\bm{y}-\bm{\Psi}\bm{a}\right)^t\bm{C}_M^{-1}\left(\bm{y}-\bm{\Psi}\bm{a}\right)+\left(\bm{a}-\bm{a}_0\right)^t\bm{C}_{aa}^{-1}\left(\bm{a}-\bm{a}_0\right)=\bm{a}^t\bm{\Psi}^t\bm{C}_M^{-1}\bm{\Psi}\bm{a} - 2\bm{a}^t\bm{\Psi}^t\bm{C}_M^{-1}\bm{y} 
$$
$$
+\bm{a}^t\bm{C}_{aa}^{-1}\bm{a} -2\bm{a}^t\bm{C}_{aa}^{-1}\bm{a}_0 + c
$$
that we can rearrange as follows,\\
$
\left(\bm{y}-\bm{\Psi}\bm{a}\right)^t\bm{C}_M^{-1}\left(\bm{y}-\bm{\Psi}\bm{a}\right)+\left(\bm{a}-\bm{a}_0\right)^t\bm{C}_{aa}^{-1}\left(\bm{a}-\bm{a}_0\right)=$\\
$$\bm{a}^t\left(\bm{\Psi}^t\bm{C}_M^{-1}\bm{\Psi}+\bm{C}^{-1}_{aa}\right)\bm{a} - 2\bm{a}^t\left(\bm{\Psi}^t\bm{C}_M^{-1}\bm{y} + \bm{C}_{aa}^{-1}\bm{a}_0\right)+c
$$
\begin{small}
$=\left[\left[\bm{\Psi}^t\bm{C}_M^{-1}\bm{\Psi}+\bm{C}^{-1}_{aa}\right]^{1/2}\bm{a} - \left[\bm{\Psi}^t\bm{C}_M^{-1}\bm{\Psi}+\bm{C}^{-1}_{aa}\right]^{-1/2}\left(\bm{\Psi}^t\bm{C}_M^{-1}\bm{y} + \bm{C}_{aa}^{-1}\bm{a}_0\right)\right]^2+c$
\end{small}
\begin{small}
$$=\left[\left(\bm{a} - \left[\bm{\Psi}^t\bm{C}_M^{-1}\bm{\Psi}+\bm{C}^{-1}_{aa}\right]^{-1}\left(\bm{\Psi}^t\bm{C}_M^{-1}\bm{y} + \bm{C}_{aa}^{-1}\bm{a}_0\right)\right)^t\left[\bm{\Psi}^t\bm{C}_M^{-1}\bm{\Psi}+\bm{C}^{-1}_{aa}\right]^{1/2}\right]^2+c
$$
\end{small}
with $[\cdot]^2=[\cdot]^t[\cdot]$. We conclude that
\begin{equation}
\label{Eq:Posterior_likelihood2}
\boxed{p(\bm{a}\vert\bm{y},\mathcal{A})=(2\pi)^{-P/2}\vert\bm{\hat{C}}_{aa}\vert^{-1/2}\exp\left(-\frac{1}{2}\left(\bm{a}-\bm{\hat{a}}\right)^t\bm{\hat{C}}_{aa}^{-1}\left(\bm{a}-\bm{\hat{a}}\right)\right)}
\end{equation}
where\\
$\bm{\hat{a}}=\left[\bm{\Psi}^t\bm{C}_M^{-1}\bm{\Psi}+\a
bm{C}^{-1}_{aa}\right]^{-1}\left(\bm{\Psi}^t\bm{C}_M^{-1}\bm{y} + \bm{C}_{aa}^{-1}\bm{a}_0\right)$ is called the Maximum A Posteriori estimate (MAP). When $\bm{C}_{aa}=\lambda\bm{I}_P$ this solution is known as the ridge regression estimator (\cite{Hoerl70TEC}).\\
$\bm{\hat{C}}_{aa}=\left[\bm{\Psi}^t\bm{C}_M^{-1}\bm{\Psi}+\bm{C}^{-1}_{aa}\right]^{-1}$ is the covariance associated to $\bm{\hat{a}}$.
\vspace{0.5cm}

\noindent\underline{N.B.}: For the same reason as previously, the posterior joint pdf should be denoted $p(\bm{a}\vert\bm{a}_0,\bm{C}_{aa},\bm{C}_M,\bm{y},\mathcal{A})=\mathcal{N}(\bm{\hat{a}},\bm{\hat{C}}_{aa})$. When $\bm{C}_M=\sigma^2_{\epsilon}\bm{I}_N$, it is not necessary to postulate $\sigma^2_{\epsilon}$ as the latter can be determined simultaneously with $(\bm{\hat{a}},\bm{\hat{C}}_{aa})$ as shown in the previous subsection (\S~\ref{Sec:OLS}).

\subsection{Laplace Prior \& LASSO}
\label{Sec:LASSO}
Let us consider now that the prior distribution is the Laplace one, namely $p_{\bm{a}\vert\mathcal{A}}=\mathcal{L}aplace(\bm{a}_0,\bm{\Lambda}_{aa})$ with $\bm{\Lambda}_{aa}={\rm diag}(\lambda_1,\dots,\lambda_P)$, $\lambda_i>0$. Then, the joint posterior distribution of the coefficients becomes,
\begin{equation}
p(\bm{a}\vert\bm{y},\mathcal{A})\propto \exp\left(-\frac{1}{2}\left(\bm{y}-\bm{\Psi}\bm{a}\right)^t\bm{C}_M^{-1}\left(\bm{y}-\bm{\Psi}\bm{a}\right)-\sgn\left(\bm{a}-\bm{a}_0\right)^t\bm{\Lambda}_{aa}\left(\bm{a}-\bm{a}_0\right)\right)
\end{equation}
performing the following transformation $\bm{b}=\bm{a}-\bm{a}_0$, yields,
\begin{equation}
\label{Eq:Posterior3}
p(\bm{b}\vert\bm{y},\mathcal{A})\propto \exp-\frac{1}{2}\left[\left(\bm{C}_M^{-1/2}\left(\bm{y}-\bm{\Psi}\bm{b}-\bm{\Psi}\bm{a}_0\right)\right)^2+2\sgn\left(\bm{b}\right)^t\bm{\Lambda}_{aa}\bm{b}\right]
\end{equation}
By developing the term in bracket, and \textbf{by assuming that $\sgn\left(\bm{b}\right)$ remains unchanged within the overall posterior solutions}, we get:
$$
\bm{b}^t\bm{\Psi}^t\bm{C}_M^{-1}\bm{\Psi}\bm{b}-2\bm{b}^t\left(\bm{\Psi}^t\bm{C}_M^{-1}(\bm{y}-\bm{a}_0)-\bm{\Lambda}_{aa}\sgn(\bm{b})\right)+c
$$
which can be factorized as follows,
$$
\left[\left(\bm{\Psi}^t\bm{C}_M^{-1}\bm{\Psi}\right)^{1/2}\bm{b}-\left(\bm{\Psi}^t\bm{C}_M^{-1}\bm{\Psi}\right)^{-1/2}\left(\bm{\Psi}^t\bm{C}_M^{-1}(\bm{y}-\bm{a}_0)-\bm{\Lambda}_{aa}\sgn(\bm{b})\right)\right]^2+c
$$
$$
\Leftrightarrow\left[\left(\bm{b}-\left(\bm{\Psi}^t\bm{C}_M^{-1}\bm{\Psi}\right)^{-1}\left(\bm{\Psi}^t\bm{C}_M^{-1}(\bm{y}-\bm{a}_0)-\bm{\Lambda}_{aa}\sgn(\bm{b})\right)\right)\left(\bm{\Psi}^t\bm{C}_M^{-1}\bm{\Psi}\right)^{1/2}\right]^2+c
$$
From this result, it can be concluded that,
\begin{equation}
\label{Eq:Posterior_likelihood3}
\boxed{p(\bm{a}\vert\bm{y},\mathcal{A})=(2\pi)^{-P/2}\vert\bm{\hat{C}}_{aa}\vert^{-1/2}\exp\left(-\frac{1}{2}\left(\bm{a}-\bm{\hat{a}}\right)^t\bm{\hat{C}}_{aa}^{-1}\left(\bm{a}-\bm{\hat{a}}\right)\right)}
\end{equation}
with the assumption that $\sgn\left(\bm{\hat{a}}-\bm{a}_0\right)$ is constant (to be checked a posteriori),\\
$\bm{\hat{a}}=\bm{a}_0+\left(\bm{\Psi}^t\bm{C}_M^{-1}\bm{\Psi}\right)^{-1}\left(\bm{\Psi}^t\bm{C}_M^{-1}(\bm{y}-\bm{a}_0)-\bm{\Lambda}_{aa}\sgn(\bm{a}-\bm{a}_0)\right)$ is the Maximum A Posteriori estimate (MAP) also called in this case \emph{Least Absolute Shrinkage and Selection Operator} (LASSO, \cite{Tibshirani96JRSS}) when $\bm{\Lambda}_{aa}=\lambda\bm{I}_P$.\\
$\bm{\hat{C}}_{aa}=\left(\bm{\Psi}^t\bm{C}_M^{-1}\bm{\Psi}\right)^{-1}$ is the covariance of $\bm{a}$.\\
\vspace{0.5cm}

\noindent\underline{N.B.}: For the same reason as previously, the posterior joint pdf should be $p(\bm{a}\vert\bm{a}_0,\bm{\Lambda}_{aa},\bm{C}_M,\bm{y},\mathcal{A})=\mathcal{N}(\bm{\hat{a}},\bm{\hat{C}}_{aa})$. Solving the problem is not trivial as one cannot guess $\sgn(\bm{a}-\bm{a}_0)$ (approximated by $\sgn(\bm{\hat{a}}-\bm{a}_0)$) before computing $\bm{\hat{a}}$. From the computational standpoint, this requires a loop over the criterion that $\sgn(\bm{\hat{a}}-\bm{a}_0)$ remains unchanged.
\section{Model selection}
\label{Sec:Model_Selection}
In this section, we discuss the model selection issue which boils down, in the present note, to finding the subset $\mathcal{A}$ such that the Bayesian model evidence, that is,
\begin{equation}
\label{Eq:BME1}
p(\bm{y}\vert\mathcal{A}) =\int_{\mathbb{R}^P}p(\bm{y},\bm{a}\vert \mathcal{A})\d\bm{a}= \int_{\mathbb{R}^P}p(\bm{y}\vert \bm{a},\mathcal{A})p(\bm{a}\vert \mathcal{A})\d\bm{a}
\end{equation}
is maximal.

There are many model selection criterion proposed in the literature, one can cite the Bayesian information criterion (BIC, \cite{Schwartz78AS}), the Akaike information criterion (AIC, \cite{Akaike73ISIT}), the Deviance information criterion (DIC, \cite{Spiegelhalter02JRSS}). In this note, we only consider the Kashyap information criterion (KIC, \cite{Kashyap82IEEE}). This criterion was derived in a Bayesian framework and is particularly suited when the input-output relationship is linear as considered in the present work (see Eq.\eqref{Eq:Model1}). In \cite{Schoniger14WRR}, it is  demonstrated that KIC usually outperforms the BIC and the AIC especially when the model is linear and the error is Gaussian. For the sake of completeness, the KIC at the MAP is derived in the next section (there is also a KIC at the MLE not discussed here).
\subsection{Kashyap information criterion}
\label{Sec:KIC}
We recall that the solution of \eqref{Eq:Posterior} is $\bm{a}\vert\bm{y},\mathcal{A}=\mathcal{N}(\hat{\bm{a}},\hat{\bm{C}}_{aa})$ under the assumption of Gaussian likelihood and prior. Let us derive the second-order Taylor series of $\ln\left(p(\bm{y},\bm{a}\vert\mathcal{A})\right)$ around the MAP, we get
\begin{equation}
\begin{split}
\ln\left(p(\bm{y},\bm{a}\vert\mathcal{A})\right) \simeq & \ln\left(p(\bm{\bm{y},\hat{a}}\vert\mathcal{A})\right)+(\bm{a}-\bm{\hat{a}})^t\underset{=0}{\underbrace{\left[\left.\frac{\partial\ln\left(p(\bm{y},\bm{a}\vert\mathcal{A})\right)}{\partial\bm{a}}\right\vert_{\bm{a}=\bm{\hat{a}}}\right]}}\\ & + \frac{1}{2}(\bm{a}-\bm{\hat{a}})^t\left[\left.\frac{\partial^2\ln\left(p(\bm{y},\bm{a}\vert\mathcal{A})\right)}{\partial^2\bm{a}}\right\vert_{\bm{a}=\bm{\hat{a}}}\right](\bm{a}-\bm{\hat{a}})
\end{split}
\end{equation}
\underline{Proof that the 2nd term is zero}: By noting that $\ln\left(p(\bm{y},\bm{a}\vert\mathcal{A})\right)=\ln\left(p(\bm{a}\vert\bm{y},\mathcal{A})\underset{\tiny\mbox{indep. of }\bm{a}}{\underbrace{p(\bm{y}\vert\mathcal{A})}}\right)$, we get $
\left.\frac{\partial\ln\left(p(\bm{y},\bm{a}\vert\mathcal{A})\right)}{\partial \bm{a}}\right\vert_{\bm{a}=\bm{\hat{a}}}=\left.\frac{\partial\ln\left(p(\bm{a}\vert\bm{y},\mathcal{A})\right)}{\partial\bm{a}}\right\vert_{\bm{a}=\bm{\hat{a}}}=0
$ by definition of the MAP.

So we come up with,
\begin{equation}
\label{Eq:Taylor}
\ln\left(p(\bm{y},\bm{a}\vert\mathcal{A})\right) \simeq \ln\left(p(\bm{y},\bm{\hat{a}}\vert\mathcal{A})\right)- \frac{1}{2}(\bm{a}-\bm{\hat{a}})^t\bm{\Sigma}^{-1}(\bm{a}-\bm{\hat{a}})
\end{equation}
with $\bm{\Sigma}^{-1}=-\left[\left.\frac{\partial^2\ln\left(p(\bm{y},\bm{a}\vert\mathcal{A})\right)}{\partial^2\bm{a}}\right\vert_{\bm{a}=\bm{\hat{a}}}\right]$. Replacing \eqref{Eq:Taylor} in \eqref{Eq:BME1} under the Laplace approximation yields,
\begin{equation}
\label{Eq:Integral}
p(\bm{y}\vert\mathcal{A}) = p(\bm{y},\bm{\hat{a}}\vert\mathcal{A})\int_{\mathbb{R}^P}\exp\left(- \frac{1}{2}(\bm{a}-\bm{\hat{a}})^t\bm{\Sigma}^{-1}(\bm{a}-\bm{\hat{a}})\right)\d\bm{a}
\end{equation}
\underline{N.B.}: The strict equality comes from the Laplace approximation that assumes that the pdf is Gaussian around the MAP.

Recalling that $p(\bm{y}\vert\mathcal{A}) =\frac{p(\bm{y},\bm{a}\vert\mathcal{A})}{p(\bm{a}\vert\bm{y},\mathcal{A})}$, it is straightforward to guess that, at the MAP ($\bm{\Sigma}=\hat{\bm{C}}_{aa}$), the integral in \eqref{Eq:Integral} is equal to,
$$
\int_{\mathbb{R}^P}\exp\left(- \frac{1}{2}(\bm{a}-\bm{\hat{a}})^t\bm{\Sigma}^{-1}(\bm{a}-\bm{\hat{a}})\right)\d\bm{a}=\frac{1}{p(\bm{\hat{a}}\vert\bm{y},\mathcal{A})}=(2\pi)^{P/2}\vert\hat{\bm{C}}_{aa}\vert^{1/2}.
$$

We conclude that,
\begin{equation}
\label{Eq:Exp_KIC}
\boxed{
p(\bm{y}\vert\mathcal{A}) = p(\bm{y},\bm{\hat{a}}\vert\mathcal{A})(2\pi)^{P/2}\vert\hat{\bm{C}}_{aa}\vert^{1/2}
}
\end{equation}

The Kashyap information criterion is defined as the deviance, namely,
\begin{equation}
KIC_{\mathcal{A}}=-2\ln\left(p(\bm{y}\vert\mathcal{A})\right) = -2\ln\left(p(\bm{y},\bm{\hat{a}}\vert\mathcal{A})\right) -P\ln(2\pi)-\ln\left(\vert\hat{\bm{C}}_{aa}\vert\right)
\end{equation}

\begin{equation}
\label{Eq:KIC}
\Leftrightarrow\boxed{
\begin{split}
KIC_{\mathcal{A}}=-2\ln\left(p(\bm{y}\vert\bm{\hat{a}},\mathcal{A})\right)-2\ln\left(p(\bm{\hat{a}}\vert\mathcal{A})\right)  -P\ln(2\pi)-\ln\left(\vert\bm{\hat{C}_{aa}}\vert\right)
\end{split}
}
\end{equation}
\underline{Conclusion}: Under the assumption that the model is linear and that the error is Gaussian, the best subset of multi-indexes $\mathcal{A}$ (or best model) is the one with the lowest KIC.

\section{Concluding remarks}
OLS (Gaussian homoscedastic likelihood+uniform prior) provides unbiased estimates of the coefficients. But it sometimes faces convergence issues, especially when the $y$-data have outliers. In that case, imposing informative prior (ridge regression or LASSO) might help overcoming this issue although providing biased estimates. Moreover, the choice of the prior's hyperparameter can be non trivial. LASSO (Laplace prior) is known to provide very sparse linear models meaning that many coefficients are set to zero and can be withdrawn from the model though. But in my experience with the KIC model selection criterion, ridge regression (Gaussian prior) performs as well as LASSO in most cases.

The formulas derived in the present document have served to derive the Bayesian sparse polynomial chaos expansion (BSPCE) algorithm in \cite{Shao17CMAME}. BSPCE takes the form of Eq.\eqref{Eq:Model1} with $\psi_{\bm{\alpha}}$ as tensor-product of univariate orthogonal polynomials. The latter has proven to be very efficient for performing global sensitivity analysis of computer model responses under the assumption that the input dataset (namely, $\bm{x}$) is sampled from independent distributions. The algorithm of BSPCE relies on a stepwise linear regression strategy. At each step, a new candidate is added to the current subset $\mathcal{A}$ and the associated KIC is assessed. If the latter is worse than the previous one, the new element is withdrawn from the subset. Notably, the $KIC_{MAP}$ is implemented with Gaussian prior (ridge regression) assigned to $\bm{a_\alpha}$ that favours low-dimensional and low-degree polynomial elements (specific choice of $\bm{C}_{aa}$).

\bibliographystyle{natbib}

\end{document}